# Sharing Begins at Home
How Continuous and Ubiquitous FAIRness Can Enhance
Research Productivity and Data Reuse

William Dempsey, University of Southern California
Ian Foster, University of Chicago and Argonne National Laboratory
Scott Fraser, University of Southern California
Carl Kesselman, University of Southern California*

**Abstract**
The broad sharing of research data is widely viewed as critical for the speed, quality, accessibility, and integrity of science. Despite increasing efforts to encourage data sharing, both the quality of shared data and the frequency of data reuse remain stubbornly low. We argue here that a significant reason for this unfortunate state of affairs is that the organization of research results in the findable, accessible, interoperable, and reusable (FAIR) form required for reuse is too often deferred to the end of a research project when preparing publications–by which time essential details are no longer accessible. Thus, we propose an approach to research informatics in which FAIR principles are applied *continuously*, from the inception of a research project and *ubiquitously*, to every data asset produced by experiment or computation. We suggest that this seemingly challenging task can be made feasible by the adoption of simple tools, such as lightweight identifiers (to ensure that every data asset is findable), packaging methods (to facilitate understanding of data contents), data access methods, and metadata organization and structuring tools (to support schema development and evolution). We use an example from experimental neuroscience to illustrate how these methods can work in practice.

**Media Summary**
Imagine that you are presented with a crowded attic full of boxes of family photos and asked to assemble an accurate photographic record of some relative's life. Most would find such a task extremely difficult. Researchers face similar challenges when attempting to comply with the data-sharing mandates that funding agencies and journals are increasingly enforcing. As they prepare a publication, they seek to assemble the data on which their results are based, but often find that these data are hard to identify, being distributed over different storage systems and labeled in ways that made sense when data was saved, but that convey little information today. Thus the data that they ultimately share are often incomplete or even inaccurate. As scientific progress depends on the ability to reproduce published research results, this situation is more than unfortunate. We suggest that this gap between intent and action is due primarily to poor process, not bad intentions, and in particular because data sharing concerns are often considered only at the end of an investigation when a publication is being prepared. We argue that the key to improved research data sharing is to adopt processes that make all data involved in a research project, in all phases, findable, accessible, interoperable, and reusable (or FAIR, in contemporary parlance). We suggest further that, just as digital cameras can simplify the family historian's life by automatically recording a time and place for every photo, so can good tools simplify the adoption of the processes required to achieve such continuous and ubiquitous FAIRness. We use an example from a multi-year neuroscience investigation to illustrate how suitable methods and tools can result in not only better data sharing at the time of publication but also greater efficiency and effectiveness in the work of the research team over the course of a project.

# 1. Introduction

We see frequent calls today for researchers to *share data*: that is, to make the data used in their scholarly research available to other investigators in ways that permit, and indeed promote, reuse. This laudable goal has spurred numerous data sharing directives, from data management plans in proposals to data publication requirements enforced by scientific journals. Frequently, these directives are inspired by the so-called *FAIR data principles*, which provide guidance for creating data that are Findable, Interoperable, Accessible, and Reusable, i.e., FAIR (Wilkinson et al., 2016).

While such directives may generate considerable electronically accessible data, it is not clear that they do much to achieve data reuse. Indeed, studies show that shared data are often erroneous or otherwise unsuited for use by others. It becomes tempting to dismiss the directives that drive their generation as mere *data sharing theater*[1]—a term used by analogy to *security theater (Schneier & Schneier, 2003)*—that is, actions that provide the feeling of doing something but achieve little. Thus, we ask: Are there approaches beyond data sharing directives that could advance the substance, as well as the form, of data sharing?

We argue in this article that the key to useful data sharing is to move beyond directives that apply only at the end of a research project; instead, we must employ approaches that integrate FAIR practices into daily research practice. By making FAIRness a *continuous and ubiquitous process*, rather than a singular act performed at the time of data publication, we will not only decrease the burdens associated with data sharing but also provide value to the researcher as the science is performed, thereby improving overall data quality.

In the following sections, we explain the need for continuous and ubiquitous FAIRness, what it means in practice, and how it can be achieved at a reasonable cost by deploying the right tools. We use an example from experimental neuroscience to illustrate how continuous and ubiquitous FAIRness can work in practice.

# 2. How the Data Lifecycle Goes off the Road

The scientific data lifecycle has long been viewed as a linear, directed path that starts with an experiment or simulation, proceeds through analysis and review phases, and ultimately terminates with a publication in an archival journal (Wing, 2019). During this final publication process, supporting data may be deposited in domain-specific repositories such as the Gene Expression Omnibus (GEO) (Clough & Barrett, 2016) or Protein Data Bank (PDB) (Berman, 2008; Berman et al., 2013) or in lightly structured repositories provided by journal publishers (Editorial, 2017) or other organizations (Blaiszik et al., 2016; Foster & Deardorff, 2017) —from where they will, ideally, be retrieved by other researchers for meta-analyses or further research studies.

---

[1] As a case in point, a resource sharing form provided by the National Institutes of Health for research funded under the BRAIN initiative provides a checklist of repositories to which data must be deposited. Yet the documentation for one of the sanctioned repositories states that instructions on how to publish a dataset are a "work in progress."

Despite the high importance assigned to the submission of data to such repositories, the data themselves are not valued adequately; they are often rife with errors. One study found that the preparation of Excel spreadsheets to be provided as supplementary material leads to erroneous gene names, due to Excel type conversions, in an astonishing 20% of papers with supplementary Excel gene lists (Ziemann et al., 2016). Interestingly, this error frequency has not improved over time (Abeysooriya et al., 2021). Such errors are not limited to supplementary material associated with publications: for example, analysis of in-situ hybridization data in the GUDMAP data repository (Harding et al., 2011) showed that 213 standard brightfield images were misidentified as being multichannel immunofluorescent images, perhaps due to improper configuration of the data acquisition software at the time of image acquisition. These are just a few examples of errors that can occur when results are assembled only at the end of a research process. Furthermore, because data submission is delayed to the end of the research, the data submitted to repositories are not organized in ways that support their use or reuse (Roche et al., 2015).

To understand how the data lifecycle can so often go off the road, we need to examine how data make their way into repositories. In many laboratories, the data to be deposited in an archival repository are assembled just prior to submission of the scientific article that they are to accompany. Much effort goes into finding, organizing, and selecting those data; many anecdotal reports suggest that this transition process is painful and error-prone, requiring hunting down data that may be distributed across notebooks, laptops, cloud storage, and local storage systems, often with inadequate metadata, and attempting to reconstruct, from ad hoc assemblages of directory structures, file names, spreadsheets, and text files, what data were used to generate the paper's results. These steps are often unsuccessful, as noted by (Tedersoo et al., 2021), who found "not enough time to search" and "couldn't find the data" to be the top two factors cited for not sharing data. The ineffectiveness of informal encodings for curation may be explained by research that shows how the consolidation of memories tends to elide episodic details over time (Lifanov et al., 2021).

It is easy to suggest that researchers should do better. But with current practices and tools, perhaps they cannot. Indeed, we argue that the reason why many data-sharing practices seem ineffective is that they are not rooted in good research practice. When data sharing occurs as an afterthought—perhaps only due to requirements imposed by a publisher or a sponsor—it provides no value to the researchers during the performance of the research and thus is often poorly executed. This situation is not a sign of carelessness but rather the inevitable consequence of attempting to bolt-on the elements required for FAIR data sharing at the end of the research process, when essential information has already been lost.

A potential solution to this problem is to view the proper organization of data generated during a scientific project as integral to the performance of the research. That is, to make FAIRness a central concern, to be addressed continuously and ubiquitously, rather than only at the time of publication. Doing so requires rethinking both work practices and data management tools, so that *all data* associated with a research project are curated and made FAIR at *every step* of the data lifecycle, from creation to publication.

## 3. Researchers are always publishing

There is a tendency to think of publication as a capstone—a singular event that occurs only once within the life cycle of an investigation. The researcher collects data, analyzes those data, organizes the results, generates figures, writes a document to describe the results, submits that document for peer review, responds to reviewer comments, and, assuming acceptance of the manuscript, eventually provides community access to selected data as an adjunct to the publication[2].

The submission of data to a mandated repository upon publication may satisfy the letter of the law when it comes to data sharing, but does it really satisfy the spirit? In considering this question, it is useful to explore the meaning of publication, the role that it plays in scientific discovery, and how its meaning and role are evolving in today's digital world.

Publications have historically filled two roles in science: claiming credit and communicating knowledge. Consider, for example, the *Philosophical Transactions*, the world's first scientific journal. On the question of credit, editor Henry Oldenberg, writing to Robert Boyle in 1664, notes: "We must be very careful as well of regist'ring the person and time of any new matter [so that] the honor of the invention will be reliably preserved to all posterity." But in the introduction to the first issue, he emphasizes communication:

> "Whereas there is nothing more necessary for promoting the improvement of Philosophical Matters, than the communicating to such, as apply their Studies and Endeavours that way, such things as are discovered or put in practice by others; it is therefore thought fit to employ the Press, as the most proper way to gratifie those, ... To the end, that such Productions being clearly and truly communicated, desires after solid and useful knowledge may be further entertained, ingenious Endeavours and Undertakings cherished, and those, addicted to and conversant in such matters, may be invited and encouraged to search, try, and find out new things, impart their knowledge to one another, and contribute what they can to the Grand design of improving Natural knowledge, and perfecting all Philosophical Arts, and Sciences." (Oldenburg, 1665)

In this regard, the *Transactions* continued what was already a long history of scientists exchanging and organizing knowledge via correspondence. For example, in the early half of the seventeenth century, French polymath Marin Mersenne developed an extensive network of correspondents, leading him to be called "The Secretary of Learned Europe" (Hauréau, 1852). These letters, which predated the existence of scientific journals, enabled collaboration among diverse scientists. More recently, computer scientist Edgar W. Dijkstra was known for his hand-written memos ("EWDs") which were photocopied and sent via post to a list of correspondents. Such less-formal "publications" are surely more frequent today, as knowledge is constantly exchanged by email, tweets, blogs, and a myriad of other methods.

---

[2] Prepublication services such as bioRxiv can modify this workflow somewhat by allowing for multiple releases of a manuscript and its data.

While the purpose of scientific publication has arguably changed little since that time, advances in technology have had major impacts on the methods of production and on what can be published. What Oldenberg termed "the Press" has expanded with the advent of online publishing and preprint servers, and a publication can now include not just prose, but also alternative media, software programs, and underlying data.

Whatever the means employed, the "improvement of Philosophical Matters" clearly benefits now, as in 1665, from frequent communication among practitioners. The perspective that a publication is an infrequent and "special" event ignores the fundamental role of the fluid exchange of ideas and data as an enabler of scientific discovery. Indeed, it may be wiser to consider any scientific communication, no matter how informal, to be a publication[3]. From this perspective, research projects involve long series of "publications," varying with the nature and number of the intended consumers: targeting, at the start of a research project, the researchers themselves (who must, after all, remember tomorrow what they did today), then fellow lab members, then members of a collaboration or consortium, and so forth. Publication in a repository is then not an event of special significance, but rather just an incremental step that further expands the range and size of the consumers of the publication, to a broad community of interest (or indeed to the public at large).

From this new perspective, data are constantly being "published." One may then ask whether publications that occur internally to a research project are any less deserving of FAIR treatment than those intended for an external repository. We take the position that this is not the case and that the same standards of curation, data management, and FAIRness must be applied for *all data* and during *all phases* of an investigation. That is, research projects must aspire to **continuous and ubiquitous FAIRness**, with the fundamentals of data management and sharing being applied unchanged over the course of a project.

An "always publishing" perspective has important ramifications. In his 1974 commencement address to Caltech (Feynman, 1974), Richard Feynman notes the centrality to science of:

> "… a kind of scientific integrity, a principle of scientific thought that corresponds to a kind of utter honesty --- a kind of leaning over backwards. For example, if you're doing an experiment, you should report everything that you think might make it invalid, not only what you think is right about it: other causes that could possibly explain your results; and things you thought of that you've eliminated by some other experiment, and how they worked --- to make sure the other fellow can tell they have been eliminated."

Operationalizing this observation in today's digital world goes well beyond keeping a good laboratory notebook or pushing final results into a community repository. Rather it speaks to profound behavioral changes so that all data associated with an investigation are collected, organized, and made shareable (e.g., FAIR). Thus, the mantra "you are always publishing" is a core aspect of the best science today.

---
[3] We note that even informal communications are increasingly considered to be citable publications. For example, the MLA style guide contains recommendations for citing a tweet.

## 4. Achieving Continuous and Ubiquitous FAIRness

Imagine the challenge that we would face if asked to document, post hoc, a box of old photos. Anybody would be hard pressed to recall when and where each photo was taken, and what people and scenes are depicted.

Today, in contrast, digital cameras can associate with each photo a time and location. Services such as Apple Photos use AI systems to identify people and objects, and automatically organize photos by time, location, participants, image quality, etc. Other tools make it easy to group photos into online albums based on topic and interest and to share those albums with family and friends, who can supplement them with additional pictures or comments. An album can easily be transferred to an online service for publication as a coffee table book, calendar, or set of coffee cups.

Modern consumer photo management tools, in effect, apply the FAIR data principles continuously and automatically as an integral part of data collection and manipulation. At every step, from data acquisition through the "publication" of an album, data are findable via an evolving set of user-provided metadata, accessible via standard HTTP protocols and authentication methods, interoperable through the use of standardized photographic file formats, and reusable via commonly shared attributes such as timestamps and location metadata[4]. Integrating data sharing technology into routine data management processes significantly improves user productivity and output quality.

A similar continuous and ubiquitous integration of FAIR data principles into routine work processes can benefit science (Madduri et al., 2019), not only reducing the potential for errors but also accelerating the creation and validation of new hypotheses. Practiced properly, continuous FAIRness streamlines the process of preparing data for publication to domain-specific repositories and increases the quality of submitted data by ensuring that each data item is linked by a clear provenance chain to the process that created it. But can we achieve these benefits without overly burdening the data producer? Our photo management example suggests that increased overhead is not a foregone conclusion if appropriate tools support practices.

## 5. From Powder Bags to Laboratory Notebooks

For an alternative to today's practice, let us consider what used to be the epitome of good research practice: the laboratory notebook, in which the disciplined scientist would, as Kanare observes (Kanare, 1985), "write with enough detail and clarity that another scientist could pick up the notebook at some time in the future, repeat the work based on the written descriptions, and make the same observations that were originally recorded."

---

[4] We note that one thing photo albums are not is *transparent*: we do not generally know whether and how a picture has been edited prior to publication.

Notebooks (whether analog or digital) are still in use today, but in this third decade of the 21st Century, the quasi-ubiquitous adoption of digital technologies, automated data collection methods, and computational processes makes the recording of "work" far more complex. The increased use of computerized methods offers both challenges and opportunities to improve on the past: if an operation is performed by a computer, then it can, in principle, be recorded automatically, in full detail, and without risk of transcription error. We need not suffer, as Arago reports (Arago, 1858) of the astronomer Le Monnier, the misfortune of failure to discover the planet Uranus, despite multiple observations being in his records:

> "If the writings were held in order, if the determinations of right ascension and declination for each day were figured in regular columns and observed, a simple glance would have shown to Le Monnier, in January 1769, that he had observed a moving heavenly body, and the name of this astronomer, instead of the name Herschel, would be found forever next to the name of one of the principal planets of the solar system. But Le Monnier's records were the picture of chaos. Bouvard showed me, at the time, that one of the observations of the planet Uranus was written on a paper bag which once upon a time held powder to powder the hair, bought at a perfumer's. Le Monnier, in reviewing his records, found that he had observed three times the so-called comet of Herschel in 1763 and 1769."

Yet one could argue that despite the exponential increases in our abilities to generate and analyze data, our laboratory practices have not, by and large, changed, with random cloud storage folders, spreadsheets, thumb drives, and servers taking the place of the scrap of paper from the perfumer.

We note that putting data in a shared location such as a lab-wide file server or a cloud-based system like Google Drive does not equate to data sharing or management. With metadata recorded in random spreadsheets or encoded in file names, it is all too easy to refer to outdated data or inadvertently share or update data in ways that do not reflect the current state of an investigation.

This situation is further complicated by the fact that while today's physical or electronic notebooks may capture process details and important contextual information, they rarely contain the actual data generated in an investigation. At best, they refer to data that must, by necessity, be located in other systems.

Researchers may account for poor record-keeping by stating that data are not ready to be properly managed. Yet as Kanare also notes:

> "Few, if any, of us working scientists write notes as carefully and completely as we should. We often lament not recording a seemingly unimportant detail that later proves crucial. Much experimental work could have been better understood, and much repetition of work avoided if only we were more attentive in our notekeeping." (Kanare, 1985).

We conclude from these observations that a perspective that "we are always publishing" is key. It requires that we employ FAIR data practices at every stage of an investigation. In so doing, we can improve transparency, avoid missed opportunities, and reduce redundancy and error. What stands in our way is that there is a gap in both tools and training between the creation of data as part of a scientific process and the publication of those data either as a standalone research artifact or as part of a publication.

Filling this gap is not just a question of tools, technology, or training in isolation. As the photographic example illustrates, the coupling of technology and workflow into an integrated socio-technical system (Baxter & Sommerville, 2011) is needed to empower communities of sharing. Such coupling would enable a transformation, streamlining the process and allowing the creation of entirely new types of outputs. Only by carefully considering the mechanisms by which researchers create, store, document, and share data daily can we reduce the burdens on the investigator, reduce practices that can result in errors and extra work, and incentivize the creation of data that ultimately can be broadly shared and reused.

## 6. Repositories Are Not Enough

The establishment and use of repositories are clearly necessary steps towards open and reproducible science. However, repositories have significant limitations as the sole mechanism for sharing data.

Repositories are often either too general or too specific. Specialized repositories such as PDB are exquisitely curated for specific data. At the other end of the spectrum, generalist repositories (Martone & Stall) such as FigShare and Dyad offer little prescriptive information about the data that they contain. Metadata are often limited to high-level descriptors. Data may be provided in arbitrarily structured directories and in different data formats. One consequence, as discussed earlier, is the potential for a proliferation of erroneous data (Abeysooriya et al., 2021; Ziemann et al., 2016). Creating a digital object identifier (DOI) to an arbitrary assemblage of data does little to advance open science and data sharing. Placing data into a repository can easily become cargo cult science: data are in the repository, so the problem is assumed to be solved.

An alternative approach would be to provide tools and methods for continuous FAIRness that impose sufficient investigation-specific structure over data throughout its lifetime, ensuring that data are always sufficiently FAIR for eventual sharing. These methods and tools may involve a combination of well-specified metadata and formats for data with established conventions (e.g., protein structures), which may be stored in appropriate repositories, along with investigation-specific metadata and representations that may be stored in other locations. The latter will often provide the context needed to understand the data placed in a repository.

The always publishing perspective can go a long way towards ensuring that data placed in generalist repositories are well structured and accurate. Rather than depositing random collections of files into a generalist repository, if the data is always FAIR, packaging it into a well-defined collection such as a BDBag (Chard et al., 2016) or RO-Crate (Soiland-Reyes et al., 2022) for placement into a generalist repository with rich and accurate metadata becomes a straightforward operation. If data are always FAIR, carving out pieces to place into both domain-specific and generalist repositories becomes a well-defined extract-translate-load process, with clear targets for extraction and loading.

## 7. It's Always a Work in Progress

The need for well-defined, rich, and up-to-data metadata is pervasive in the FAIR guidelines (Wilkinson et al., 2016). In the absence of tool support, researchers are typically on their own with respect to these requirements. Thus, we often see metadata encoded in structured file naming conventions, which are hard to maintain and limit discovery and reuse. More sophisticated users may create local databases, but such databases typically require knowledge of query languages and are difficult to set up and maintain. Increased use of embedded databases such as SQLite, along with cloud-hosted structured and unstructured databases (e.g., NoSQL), can simplify the deployment and operation of metadata stores but still leave the user the tasks of creating up-to-data metadata descriptions and data curation.

An additional challenge is that an inevitable consequence of research is that both metadata schema and their associated values tend to change over time as a research project evolves. Indeed, it is almost axiomatic that metadata schema, relationships, and values will evolve during research, as new discoveries drive new insights. For example, researchers may want to coin a new term to describe a new observation, document a new type of entity, or record more detail than previously, as they understand more about the research question being studied. Tools designed to maintain FAIRness over the lifecycle of an investigation must support such metadata evolution.

In keeping with the theme of this paper, we note that change is driven not only by changing demands over time *within* a community (continuity) but also by changes in the *extent* of the community (ubiquity). For new scientific results, almost by definition, widely agreed-upon community ontologies or controlled vocabularies may not yet be available and will only emerge over time (Noy & Klein, 2004). However, there is a clear benefit to defining and reusing terms consistently even within more informal publications, eventually evolving those initial terms into broadly accepted community models and vocabularies. This approach mirrors that adopted by the Dublin Core, which defines a data "pidgin" language that may be enhanced in domain-specific ways to create a "creole." Over time, the creole evolves, enabling effective publication and communication within and across communities and creating the foundation for more formalized descriptions to be developed by slower-moving community processes for broader dissemination and reuse (Baker, 1998).

A good example of this type of language evolution can be found in gene naming conventions. Gene names are commonly assigned by the discoverer and often reflect the function of that gene in the specific scientific domain in which they work. Some communities are more fanciful than others in the names that they choose. For example, the fruit fly community had a propensity for names that allude to a gene's impact on a fly's physical development, resulting in names like Indy (for "I'm not dead yet"), Lush, Groucho, and Lunatic Fringe. These names were commonly used in the fruit fly community; that is, they were a creole. However, as connections to human disease were discovered, the Human Genome Organization (HUGO), concerned that it would be problematic to tell a patient that, for example, they suffered from a "Lunatic Fringe based disorder," developed new names, such as LFNG for what was previously Lunatic Fringe (Hopkin, 2006)

We thus need methods and tools that allow schema and metadata to adapt readily to changes over the lifecycle of an investigation. In the database community, solving the *schema evolution problem* generally is challenging (Stonebraker et al., 2016). For the practicing researcher, less comprehensive solutions can still be helpful. For example, it will often be good practice to organize metadata from the start to include well-defined points for augmenting term sets and to build descriptions of core entities around the idea that they may be extended by using standard techniques such as extension tables in a database. Failure to appreciate and apply even basic techniques can have significant consequences. For example, in one situation experienced by the authors, a shared database of X-Ray crystallography experiments designed by a team member used more than ten different terms to indicate that an experiment had been completed.

Ultimately, the best solution may be to create tools designed specifically to enable researchers (or informatics-savvy members of a research team) to modify data representations incrementally. These tools can be used to maintain metadata descriptions to be representative of the current structure of a research project (Schuler et al., 2020).

## 8. Tools Can Help

We have proposed that researchers need to apply FAIR principles to all of their data, not just their published data, and used the example of consumer photography to illustrate how, with the appropriate tools, this process need not be onerous. But research is, of course, different from consumer photography. What specifically do we need tools to do?

The FAIR principles provide a good starting point for analysis. These principles require, among other things, naming—and associating descriptive metadata with, and recording provenance for—every digital object (Wilkinson et al., 2016). Under continuous and ubiquitous FAIRness, we apply the FAIR principles more expansively in both **space** (to all data created during a research project, not just the data that are published) and **time** (at every stage of the project, not just at the time that a paper is accepted). This more expansive application then demands tools that **automate important elements of the overall process** to reduce both opportunities for human error and incentives for shortcuts that would reduce data FAIRness. Meanwhile, a socio-technical perspective (Baxter & Sommerville, 2011) on continuous and ubiquitous FAIRness suggests that in proposing new tools, we must take a user-centric approach in which changes in practice are balanced with ease of use and deliver immediate as well as long term benefit.

One factor working in our favor is that many researchers already use proprietary and open-source tools such as MatLab, Jupyter notebooks, R-Studio, Excel, and standardized Python and R libraries to organize elements of their work processes. However, these tools are commonly used in idiosyncratic and unstructured ways that tend to focus on code and computing rather than on the management of the associated data. The challenge, therefore, is how to complement these existing tools in ways that enhance FAIRness while minimizing impacts on existing workflows.

A first useful step is often to impose more structure on how tools are used. For example, Sandve et al. (Sandve et al., 2013) make ten suggestions on improving reproducibility via the structured use of existing tools, including GitHub for code repositories and programs rather than manual operations for data manipulation. Environments for tracking data provenance, such as WholeTale (Brinckman et al., 2019), can also address a number of the steps identified in this paper. However, in general, tools that focus on effective data management by FAIR principles are lacking. In recognition of these gaps, in our own work, we have explored what types of tools could be readily integrated into the daily work of a researcher that would promote the continuous and ubiquitous creation of FAIR data.

**Create and resolve lightweight identifiers**. It is widely agreed that persistent globally unique identifiers are essential to FAIR data production (Chodacki et al., 2020). Yet, in current practice, such identifiers are typically only generated at the time of data publication. Our focus on continuous and ubiquitous FAIRness demands that we have such identifiers for any possible data artifact. In addition, we need to be able to create identifiers quickly and easily for many different data, at varying levels of resolution (from a single file to larger collections) and that record a binding to a specific sequence of bits so that modifications can be detected. To address these needs, we have developed a set of open source tools and services for creating minimal identifiers, or Minids (Chard et al., 2016): identifiers that are sufficiently simple to make their creation and use trivial, while still having enough substance to make data easily findable and accessible. Minids can be created in the thousands (or millions), resolved quickly, and upgraded to DOIs when data are published.

Minids can be created via command-line tools or via programmatic APIs through programming languages such as Python. Consistent with their name, Minids encode very little information: just the author, creation date, checksum associated with the data, and one or more URLs indicating where to find the data. The Handle system is used to resolve the identifiers, much as Digital Object Identifiers (DOI) are resolved (Paskin, 2010). By providing the ability to easily integrate the creation and resolution of robust identifiers bound to specific data content into the scripts and workflows used to generate data, we can enhance the FAIRness of those data throughout their lifecycle.

**Create and manipulate data aggregations**. Another frequent source of friction in research workflows results from the need to create, operate on, and share data aggregations. For example, the input to a biomedical analytic pipeline can be a dataset consisting of thousands of images and genome sequences assembled from diverse repositories. FAIR sharing requires that the contents of such datasets be described in a concise and unambiguous form. Ad hoc approaches like directories are commonly used for this purpose but assume that all data reside in a single location, requiring costly data marshaling, and encourage errors of omission and commission because dataset members are not explicitly specified. Here, we have found it helpful to leverage a technology called BDBags (for Big Data Bags), packages defined according to the BagIt specification that conforms to the BDBag and Bagit/Research Object (RO) profiles (Chard et al., 2016).

In brief, a BDBag provides a simple and robust method for describing data collections that overcome the limitations of directories, make it easy to associate checksums with data, and are naturally combined with Minids as identifiers and with Research Objects (Chard et al., 2016) for providing data descriptions. While other packaging formats exist (e.g., RO Crates (Soiland-Reyes et al., 2022), GZIP (Deutsch, 1996)), we choose to work for Bagit because a) it has been formally defined via RFC 8493 (Madden et al., 2018) b) it completely defines the container contents via checksums and manifest, c) it has a flexible and extensible mechanism for including metadata, and d) it includes the ability to refer to file contents that are not directly included in a bag by reference (called a holey bag), enabling compact representations of potentially large datasets. Bags are easily created by using tools familiar to scientists who have created ZIP, TAR, or other file archives. The current BDBag tooling has a plugin architecture and can assemble the contents of a holey bag referred to by URL, DOI, Minid, or other naming schemes. Metadata in a BDBag can be specified by using the research object schema, key/value pairs, or other more structured mechanisms, such as table schema (Walsh & Pollock, 2021). Like Minids, BDBags were designed to provide basic, easy-to-use tools to enable FAIR data management with minimal intrusion into the daily workflow of a scientist.

Because Minids and BDBags are open source and are supported by simple Python libraries and command-line tools, they are readily used from within existing programming environments, such as Python programs and Jupyter notebooks, and easily integrated into a range of applications and workflow tools, enhancing our ability to track data products and ultimately to share FAIR data (Chard et al., 2016; Madduri et al., 2019).

**Streamline Publication.** Automation of repeated processes is essential to reduce the risk of both human error and, equally problematic, shortcuts that reduce data FAIRness; it also has the benefit of making researchers more productive. A repeated process might involve, for example, data capture from an instrument; transfer to persistent storage; preparation of metadata; packaging with metadata in a structured container; quality control; DOI generation; and cataloging. Typically, each step is straightforward, but if the process is performed manually, individual steps can easily be misconfigured or omitted due to time pressure, especially when complicated by concerns such as authentication and error handling. For automation, we increasingly make use of Globus Flows, a technology for capturing such series of steps (Ananthakrishnan et al., 2018).

**Manage Data Assets.** We noted how tools specifically targeted towards the management of pictures have transformed how photographs are created and used. Photographic management is an example of a class of systems known as *asset management platforms*, software tools specifically designed to support the management of data in unstructured work environments. These systems provide user-centered interfaces for managing relationships between files (e.g., digital assets) and metadata models for purposes of discovery, organization, analysis, and sharing. Asset management tools designed to support institutional data sharing goals, such as DSPACE (Smith et al., 2003; Tansley et al., 2005) and Google Goods (Halevy et al., 2016), have met with success. However, these platforms are rarely integrated into the daily work of research projects. The authors have explored how to create asset management systems that are more readily integrated into the daily work of a researcher in service of continuous FAIRness.

Building on the ideas of data-centric workflows and asset management systems, the Deriva ecosystem (Bugacov et al., 2017; Schuler et al., 2016) is specifically designed to support the integration of FAIR data practices into the ongoing research investigations. Deriva is an open-source, general-purpose platform that has been configured and deployed into a diverse range of collaborative and investigative use cases, including protein structure determination (Vallat et al., 2021), developmental biology (Harding et al., 2011), craniofacial-dysmorphia (Samuels et al., 2020), cellular modeling and optimization (Deng et al., 2019), to name a few examples.

Deriva provides basic services for FAIRness (e.g., descriptive metadata, the pervasive use of permanent identifiers, standard open-access methods, and fine-grain access control) while simultaneously adapting its underlying data model (e.g., metadata) to support the continuous evolution of an investigation over time.

A data catalog service is used to track assets via the descriptive metadata required to organize and discover those assets (Czajkowski et al., 2018). Assets may be digital objects (e.g., data files, protocol specifications) or physical objects (e.g., mice, biosamples). The catalog tracks references to these assets and descriptive metadata—contextual information on how assets are generated, relationships between assets, and other descriptive concepts (e.g., species, genes, or antibodies).

The catalog structures the asset references and descriptive content by using an Entity Relationship Model (ERM), loosely referred to as the data model. The catalog supports the user-driven creation, editing, query, exchange, and storage of metadata about both individual data items (e.g., files) and collections of data items (e.g., a group of sequencing files that constitute a controlled experiment or a set of images and annotations that capture gene expression observations). Every element in the catalog is automatically assigned a persistent ID, and every data and metadata element is versioned. In addition to a faceted interface, programmatic APIs are defined for all functions, enabling easy integration into data processing environments such as Jupyter notebooks.

As an investigation proceeds, the structure of the data being managed often changes as protocols and approach are refined. These changes in structure can be a barrier to adopting FAIR practices as data is generated. To address this problem, Deriva was designed not only to adapt easily to different application domains but also to evolve incrementally over the duration of a scientific investigation. To this end, it provides tools specifically designed to enable domain scientists to modify their data model without requiring the services of a database administrator (Schuler & Kesselman, 2021).

## 9. A "Memorable" Example – Applying Ubiquitous FAIRness in Neuroscience

We illustrate the benefits of continuous and ubiquitous FAIRness by describing a multiyear investigation involving three of the authors in which these principles were applied. This investigation had as its overarching goal the development of a new technique for monitoring changes in sub-micron scale connections between the functional storage units of the brain, neuronal synapses (Fig. 1A), within a living organism, to study how the overall structure of these connections changes with learning (Dempsey et al., 2022b). This example leverages the Deriva scientific asset management platform described in the previous section, which it used to refine the use case-specific data model over time as the study progressed.

The idea that memories are encoded in the brain by changes in the strength of synaptic connections was proposed by Santiago Ramón y Cajal in 1894 (Jones, 1994). Today, neuroscientists broadly accept this perspective, and since the early 1970s, studies have examined and characterized the cellular structure of neurons and their connections in organotypic brain slices (Humpel, 2018). Yet synapse-level changes that occur within intact organisms during the acquisition of conceptual knowledge (e.g., aversion learning) remain poorly understood. Since synapses are considered the key functional elements of a neuron, studying their organization, connecting partners, and structure in space over time should give insight into how an abstract concept like a memory may be encoded by changes in the physical architecture of the brain.

To approach this complex issue, a set of tools that enable direct observation and spatial mapping of reporter-labeled synapses without damaging the animal were developed and applied (Dempsey et al., 2022b). The synapses themselves are labeled by using a genetically encoded reporter that targets postsynaptic densities on one side of the synaptic cleft (Fig 1B) (Gross et al., 2013). Using these tools, the microstructure of the pallium of larval zebrafish that occurs during memory formation resulting from exposure to a novel classical conditioning paradigm was tracked and recorded. As expected at the onset, this five-year study produced terabytes of data and downstream analyses. Consequently, the research team decided to manage their data and analytical pipeline with a data management system designed to support continuous and ubiquitous FAIRness, described in more detail below.

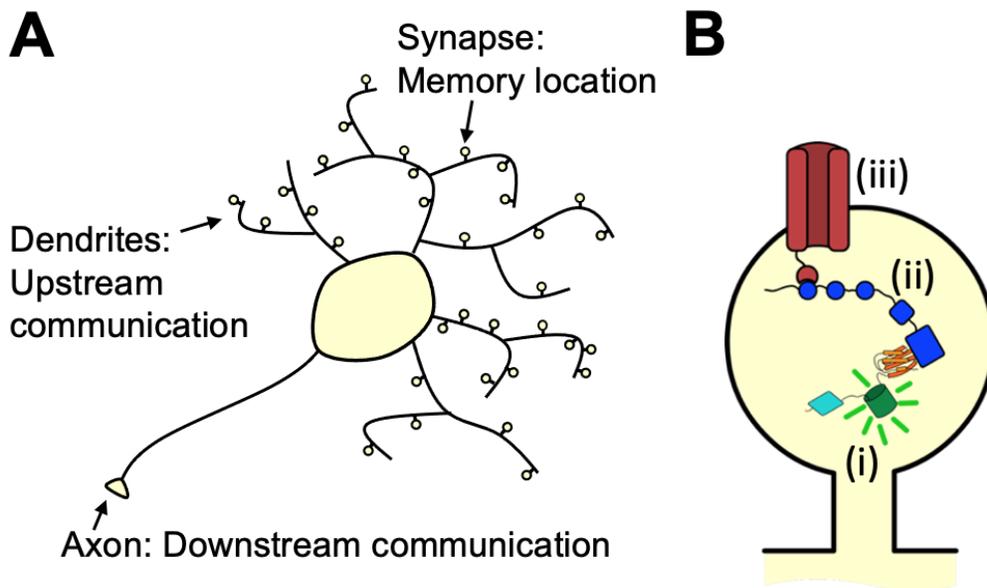

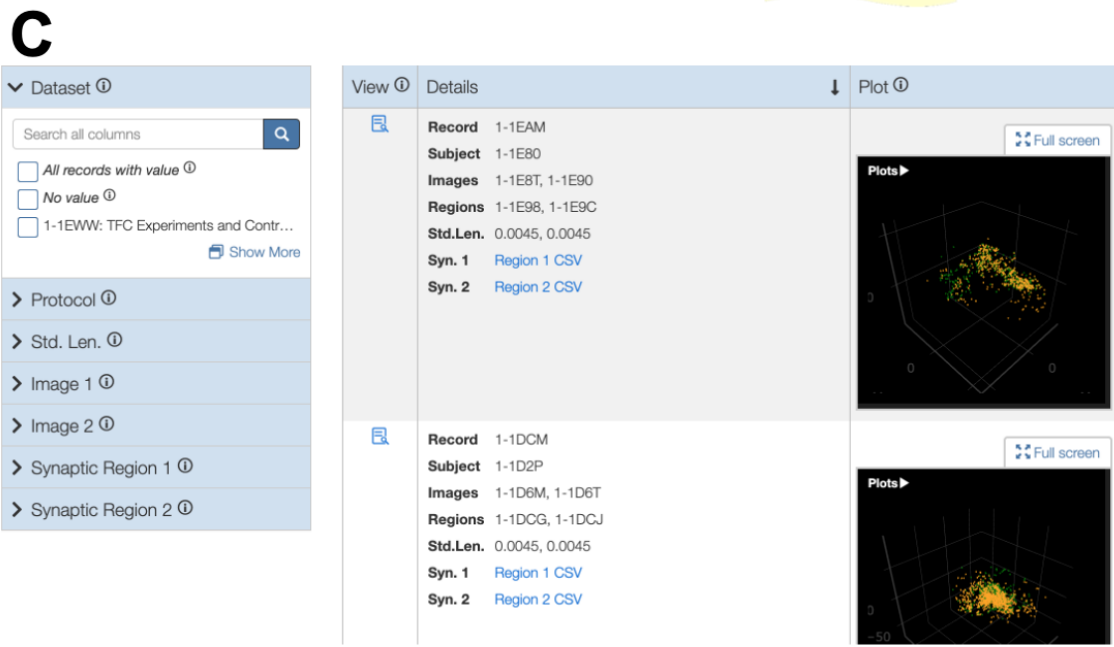

**Fig. 1**: Continuous and ubiquitous FAIR data management as applied in a neuroscience application. **A**, Neurons are the functional units of the brain, involved in various processes, including locomotion, learning, and consciousness. Canonically, neurons receive inputs into postsynaptic terminals ("synapses") within dendrite arbors. Signal integration occurs in the cell body, and downstream communication is facilitated by the axon, which terminates in a presynaptic terminal that contacts downstream postsynaptic terminals. **B**, Synapses are labeled with (i) a fluorescent reporter that binds (ii) a key synaptic scaffolding element (PSD-95), responsible for anchoring important (iii) receptors in the membrane (Gross et al., 2013). **C**, Exemplary portion of a searchable table within the data management system. After experimentation, fluorescence emitted by these synapse labels and recorded during the experimental protocol is analyzed in three dimensions. Positional and intensity information from each synapse is maintained within the data management system in an intuitive file structure, along with downstream results as well as key metadata such as timepoint, experimental procedure, key reagents, and animal life stage.

This data management-conscious approach facilitated close coupling between a multidisciplinary team of researchers. Novel hardware and software tools were developed concurrently, including an optimized selective plane illumination (SPIM) microscope, in vivo genetically encoded reporters, a behavior training apparatus, new software and methods for identifying synapses in low signal-to-noise microscope images, and analytical techniques for understanding large-scale changes in synaptic structure. The refinement of these new methods over the course of the investigation ultimately involved hundreds of experiments.

Crucially, the study was conducted such that all data produced by every experiment were captured in a standard format, preserved, described, assigned a persistent identifier (e.g., doi:10.25551/1/1-1F6W), and cataloged to permit subsequent discovery and retrieval. The catalog also linked each data record to the instrument, experimental protocol, software, etc., used to generate it (Fig. 1C). Thus, at any point in time, the researchers could readily determine, for example, what experiments had been conducted, which protocol was used to generate a specific data item, and how data from one instrument compared with data from another. In other words, *all* data, from the first experiment to the published results, were *always* FAIR. The key to achieving this continuous and ubiquitous FAIRness was our development, concurrently with the experimental protocols outlined above, of a supporting scientific data management system based on the Deriva platform.

We next elaborate on how the data management system is used on a day-to-day basis. A researcher involved in the study begins a particular experiment by first writing a descriptive protocol in the data management system (Fig. 2A). Next, they enter each experimental subject (in this case, larval zebrafish) as a unique entity into the system under this or another protocol, along with relevant contextualizing and descriptive metadata (e.g., date of experiment, protocol, imaging method used, behavioral training used). Different stages of the experimental pipeline are linked together with a modular and readily modifiable data model (Fig. 2B). Specifically, the starting point of the experiment, the zebrafish subject, is linked to all downstream data, including behavioral training and brain imaging. The data model flow diagram is accessible on the data management system home page. Each node is represented as a link that directs users to the relevant tables in the repository. New subject entries and data for all currently available protocols can be easily input by a researcher with minimal effort. New nodes can be entered into the model in an evolutionary manner without disrupting current datasets. One of the advantages of the Deriva model is that the interfaces for search and data entry adapt to the underlying data model.

Day-to-day maintenance and enhancements over time on the data management system were handled by a collaborative team consisting of a research software engineer (working at ~50% capacity) and one domain-specific researcher (up to 50% of this person's effort), providing feedback and ensuring compliance with other researchers on the team. In the span of five years, we estimate the total cost of maintaining the system to be less than USD 250K in a project with an overall budget of over USD 9.7M. The primary team for the study consisted of five individuals working full time and eight others engaged in specific portions of the study over time. When also counting research staff, interns, and undergraduate researchers who had some role in the study, more than 25 people became reliant on the data management system over the course of five years. Thus, only a small portion of the overall cost of the study went into the development and maintenance of the data management system, especially relative to the cost of personnel and materials. Having faster, more reliable collection, analysis, and organization of data may outweigh the modest up-front cost of such a system, but the added benefits of improved long-term reproducibility and transparency made the choice of maintaining this system a top priority regardless of price.

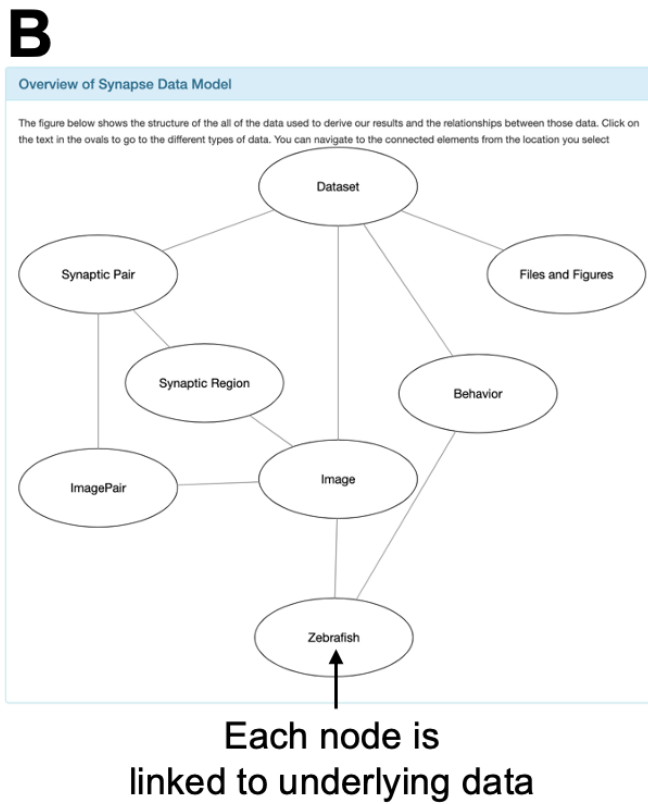

**Fig. 2**: Data entry, display, and data model in the data management system used in the synapse study. **A**, Intuitive web interface for setting up a new protocol definition. Researchers can easily add new protocols for each relevant phase of the study, which can then be defined to have "protocol steps" linked in a parent-child fashion. Data from each zebrafish subject are assigned a particular protocol before being entered into the system. **B**, All major portions of the data model above the level of experimental protocols and resources, including the entries of the subjects themselves ("Zebrafish") and the combination of all data and analyses ("Dataset"). Each node in the model can be clicked to link to the relevant data for that particular node. **C**, At the end of each study, a list of manuscript figures and all study data is available as hyperlinks into the data management system. The model and manuscript-related data can all be accessed on the front page of the Web interface: synapse.isrd.isi.edu.

The Deriva-based data management system assured that all data, from the first experiment to the published results, were FAIR. A welcome consequence of this data-conscious approach is an uncommonly high degree of transparency and reproducibility. Each figure is directly supported by the data that went into its creation, and all raw microscope data, analyzed data, human interpretations of these data, analysis programs, and essential research resources are available in the same structured, queryable, and downloadable form that was used to conduct the study. *Making data management a top priority from the start ensures that data sharing is no longer an afterthought but is instead an implicit result of the study*.

**Additional benefits and insight gained from the data management system**. In the early stages of the investigation, when data management needs were first discussed, the researchers involved in the study knew that they could quickly become overwhelmed with the complexity and diversity of the forthcoming data. They were concerned about the potential for errors being introduced during data collection, analysis, and presentation. The synapse project team addressed this concern by making the data management system a socio-technical ecosystem in which interactions with the data management platform became an integral part of the daily scientific work process. To this end, the Deriva platform engineers took the time to beta test model components iteratively to see what features would result in the highest level of compliance and ease of use for the researchers, understanding that the system would only operate efficiently if the users were excited to use it. An advantage of the Deriva tools is that they can readily adapt to changes in data model over time; hence the data management system always reflected the current state of the investigation, making sure to reduce the effort required to enter data into the system wherever possible. This user-centered approach and focus on matching tools with daily work processes resulted in all investigators readily adopting the system.

The data management solution was comprehensive. Even the most mundane metadata was entered into the system, such as the transgenic status of each zebrafish subject and the Research Resource Identifier (RRID) (Hsu et al., 2019) locations of key software and reagents used during experimentation. Users could define controlled vocabulary elements to minimize the chance of user error during input. Controlled vocabulary was especially important when inputting the genetic signature of each transgene. Gene sequence definitions were often similar in name (e.g., promoters, regulatory elements, reporter elements), so making sure to input the proper sequence was critical. Researchers could easily add controlled vocabulary elements in relevant tables within the catalog. Additionally, each device (e.g., microscopes) built for the experiment was uniquely identified with a controlled vocabulary, with its configuration recorded. As mentioned previously, all experimental protocols were entered into the system as they were designed (Fig. 2A). Imaging data from all microscope observations were uploaded into the system immediately after production. Metadata from these images were automatically entered into the catalog since the microscope data was stored in a data-rich format, such as the .ome.tiff data structure (Goldberg et al., 2005).

All software, maps of synapse position and intensity over time, and analytic output were also registered, each with descriptive metadata and unique persistent ID. Access controls were configured according to a researcher's role within the experiment. The Deriva platform provided versioning on all metadata values and checksums and versioning on all experimental data. In addition, we decided early on that write privileges should be given only to the primary researchers within the experiment; to reduce the risk of entries being accidentally deleted, created, or mislabeled. Interns, undergraduates, and early-stage users of the data management system were granted read-only access to all data and write access only to the data that they were responsible for. Since only a few researchers had direct write access to the data management system, any mistaken entries could easily be resolved. We went one step further for our short-term intern and undergraduate researchers, developing a bespoke, GUI-driven toolbox that allowed them to download raw data and analyze those data with the same interface. Uploading processed data required simply pressing a button at the end of their analytical pipeline.

As expected in a multi-year endeavor, improved methods to analyze and display data were developed over time. This eventuality was anticipated from the beginning. For example, an improved algorithm was implemented to analyze behavioral data several years after the study began. The integrated nature of the data management system allowed each of the 200+ behavioral datasets to be readily identified and re-analyzed automatically with this new algorithm, with new figures displayed within the repository immediately after, without the need for the researchers to intervene manually.

The comprehensive data and metadata collection supported by the data management system had additional unexpected benefits that became apparent through the course of the study. The collection and subsequent inspection of "failed" data allowed the researchers to optimize their experimental equipment and procedures. For example, analysis of "failed" data traces from one set of experiments led to the identification of the failure mode based solely on the recorded metadata in the data management system. It turned out that these failed datasets all came from the same microscope. Since multiple researchers used each of the different microscopes, it would have been challenging to identify which microscope was malfunctioning if they had not diligently recorded each result immediately after each three-hour-long experimental campaign. Once the underlying issue was uncovered, the researchers were able to repair this device quickly and continue with experiments at maximum efficiency.

Another set of "failed" experiments allowed the researchers to pinpoint a small change that was made in setting up the imaging environment. Even with the advantage of the data management system's detailed catalog, several weeks were lost due to this seemingly minor deviation from the standard imaging protocol. Without appropriate data management, this small change might never have been discovered. The use of the data management system thus avoided what would likely have otherwise been a significant negative—and costly—effect on the study.

From the perspective of archival publication, the normally tedious process of preparing figures to submit to peer-reviewed journals was radically simplified. Once the panels for a specific figure were arranged, data supporting that figure were simply referenced via a DOI that was already automatically created for each data product in our experimental repository. The researchers simply needed to include that DOI in the figure caption. Importantly, no additional work was required to collect the data referenced by that DOI for inclusion in a domain-specific repository. Significantly, as the contents of the paper and associated figures evolved during writing and review, the associated data was referenced with no additional work required by the authors. Ultimately, all figure links and links to all study data were placed on a simple web page associated with the publication for easy navigation (Fig. 2C) (Dempsey et al., 2022a).

**Closing thoughts**. The application of continuous and ubiquitous FAIRness in this neuroscience study did not impose significant overhead on the research process. We estimate costs associated solely with the configuration and maintenance of the platform for data management to be in the range of 10-15% of the overall project budget. However, in assessing the overall costs associated with applying continuous and ubiquitous FAIRness, we should also consider the improved efficiency of the research team due to the ease with which they could identify and correct various errors, implement improved algorithms for data analysis, and prepare data for peer-review submission. The transition of data to domain-specific repositories also becomes trivial.

Ultimately, the purpose of FAIRness is to encourage reuse. An objective determination of whether the continuous FAIRness methodology supported by Deriva enhances data reuse is difficult given that we have no controlled experiment with which to compare. However, an subjective assessment based on comparison with current standard practice suggests that without Deriva, metadata would be extremely limited, as it would have been added only *after* data were collected, and would probably comprise just high-level textual descriptions of subject and keywords. And while the keywords might have been selected from controlled vocabularies such as schema.org, they would not address details internal to datasets, and thus discovery within datasets would have been all but impossible. We conclude from this thought experiment that the data generated via Deriva is FAIRer and thus more reusable. We can see indications of such positive outcomes in other domains, such as craniofacial research (Schuler et al., 2022) which has adopted Deriva to achieve some aspects of continuous FAIRness, although in a more limited implementation. In those other examples, we have seen significant documented growth in both data contributions and data reuse since adopting a Deriva-based approach to FAIRness.

Thus, this use case provides clear evidence that straightforward adoption of a strict yet adaptable data management solution is possible—even when dealing with sophisticated biological applications—and can deliver multiple advantages for researchers willing to adopt a FAIRness-centric mindset wholeheartedly.

## 10. Knowing Enough To Be Dangerous

A significant contributor to the success of the neuroscience use case was the luxury of having a trained computer scientist participate in the research team. This was possible because this role was written into the initial project proposal and was a core part of the research budget. Unfortunately, this situation is not common. While researchers will often budget for cores and personnel for activities such as bioinformatics, biostatics, sequencing, and imaging, in the authors' experience, it is rare to see data management, FAIR or otherwise, called out as an explicit activity in research proposals. More typical is that funding agency requirements for a data sharing plan is met by the allocation of responsibility for executing that plan to the most "IT literate" graduate student in the research group, who knows enough to be dangerous–but typically lacks any formal training in data management, software engineering, or other necessary background skills, with cargo-cult science being all too often the result (Gibbs et al., 2017).

We observe, however, hopeful trends. The increased focus on reproducibility and open data by funding agencies creates an incentive to change (Kozlov, 2022). The growing importance of data science and the creation of new multidisciplinary programs oriented towards training scientists in best practices for data management and analysis are increasing the number of researchers who understand data sharing issues and methods. Finally, the emergence of research software engineering as a career path (Katz et al., 2021) is creating a workforce that knows enough to be helpful. These trends lead the authors to conclude that the approach to achieving continuous FAIRness described here can be replicated more broadly.

A reader may ask: do the modest overheads observed in the neuroscience use case scale with project size, so that for example smaller research teams can also apply the methodology? We have empirical evidence that suggests that this is the case. The methodology and Deriva tools used in the neuroscience study have also been applied to projects involving just a few graduate students and limited faculty time, in topics such as numerical optimization (Deng et al., 2019) and modeling cell function (Singla et al., 2018), with similarly good results. On reflection, these successes should not be surprising, given that Deriva and the Globus tools described above are cloud-hosted services with minimal deployment and operation costs and that software tools such as BDBag are provided via well-established distribution platforms such as PyPi (*The Python Packaging Index*)—and that, as noted earlier, compliance with FAIR guidelines tends to lead to the efficiency gains.

Given the increasing centrality of data to scientific discovery and our experiences to date, we assert that with appropriate tools and training, the methods we describe in this paper can be used effectively by even small, single investigator awards. In point of fact, these are precisely the types of investigations that cannot afford *not* to use these methods, as they typically do not have the wherewithal to develop bespoke data management tools.

## 11. Summary

Studies continue to show shockingly high levels of irreproducible research. While many factors contribute to this crisis, open data accessible from repositories remains a significant issue, as evidenced by a recent paper reporting data availability in less than two percent of the cohort of papers examined (Errington et al., 2021). We have argued here that more tightly integrating elements of data sharing and reproducibility into the generation of research results can improve the quality of those results and enhance the ability for others to use and reproduce them. We used a detailed example, drawn from neuroscience, to show that these improvements can be realized by non-experts in dynamic, data-driven research projects if they are integrated into the daily practice of the research team in such a way as to provide direct short-term benefits and adopted as an essential component of the research process.

## Disclosure Statement
The authors acknowledge support from the NIH, DOE, and NSF.

## Acknowledgments

The authors acknowledge the significant contributions of the synapse research team to the development of the use case reported in this paper. Team members include:
Zhuowei Du, Anna Nadtochiy, Colton D Smith, Karl Czajkowski, Andrey Andreev, Serina Applebaum, Thai Truong, and Don B. Arnold. Karl Czajkowski was primarily responsible for creating and maintaining the Deriva based data management platform for the synapse experiments. Mike D'Arcy developed specialized applications for curating synapse maps.

## Data Repository/Code

All of the data associated with the synaptic measurement use case can be found at: https://doi.org/10.25551/1/1-1JR0.

The GitHub repositories for BDBags and Minids can be found at https://fair-research.org. Source code repositories for Deriva can be found at: https://github.com/informatics-isi-edu.

## Supplementary Files (optional)

*Supplementary files should be submitted as separate files and appear as links in published version.*

## References

Abeysooriya, M., Soria, M., Kasu, M. S., & Ziemann, M. (2021). Gene name errors: Lessons not learned. *PLoS Comput Biol*, *17*(7), e1008984. https://doi.org/10.1371/journal.pcbi.1008984

Ananthakrishnan, R., Blaiszik, B., Chard, K., Chard, R., McCollam, B., Pruyne, J., Rosen, S., Tuecke, S., & Foster, I. (2018). Globus platform services for data publication. In *Proceedings of the Practice and Experience on Advanced Research Computing* (pp. 1-7).

Arago, F. (1858). *Popular astronomy* (Vol. 2). Longman, Brown, Green, and Longmans.

Baker, T. (1998). Languages for Dublin Core. *D-lib Magazine*.

Baxter, G., & Sommerville, I. (2011). Socio-technical systems: From design methods to systems engineering. *Interacting with computers*, *23*(1), 4-17.

Berman, H. M. (2008). The Protein Data Bank: a historical perspective. *Acta Crystallogr A*, *64*(Pt 1), 88-95. https://doi.org/10.1107/S0108767307035623

Berman, H. M., Kleywegt, G. J., Nakamura, H., & Markley, J. L. (2013). How community has shaped the Protein Data Bank. *Structure*, *21*(9), 1485-1491. https://doi.org/10.1016/j.str.2013.07.010

Blaiszik, B., Chard, K., Pruyne, J., Ananthakrishnan, R., Tuecke, S., & Foster, I. (2016). The Materials Data Facility: Data Services to Advance Materials Science Research. *JOM*, *68*(8), 2045-2052. https://doi.org/10.1007/s11837-016-2001-3

Brinckman, A., Chard, K., Gaffney, N., Hategan, M., Jones, M. B., Kowalik, K., Kulasekaran, S., Ludäscher, B., Mecum, B. D., & Nabrzyski, J. (2019). Computing environments for reproducibility: Capturing the "Whole Tale". *Future Generation Computer Systems*, *94*, 854-867.

Bugacov, A., Czajkowski, K., Kesselman, C., Kumar, A., Schuler, R., & Tangmunarunkit, H. (2017). Experiences with DERIVA: An asset management platform for accelerating eScience. 2017 IEEE 13th International Conference on e-Science (e-Science),

Chard, K., D'Arcy, M., Heavner, B., Foster, I., Kesselman, C., Madduri, R., Rodriguez, A., Soiland-Reyes, S., Goble, C., & Clark, K. (2016). I'll take that to go: Big data bags and minimal identifiers for exchange of large, complex datasets. 2016 Ieee international conference on big data (big data),

Chodacki, J., Hudson-Vitale, C., Meyers, N., Muilenburg, J., Praetzellis, M., Redd, K., Ruttenberg, J., Steen, K., Cutcher-Gershenfeld, J., & Gould, M. (2020). *Implementing Effective Data Practices: Stakeholder Recommendations for Collaborative Research Support.*

Clough, E., & Barrett, T. (2016). The gene expression omnibus database. In *Statistical genomics* (pp. 93-110). Springer.

Czajkowski, K., Kesselman, C., Schuler, R., & Tangmunarunkit, H. (2018). Ermrest: a web service for collaborative data management. Proceedings of the 30th International Conference on Scientific and Statistical Database Management,

Dempsey, W. P., Du, Z., Nadtochiy, A., Smith, C. D., Czajkowski, K., Andreev, A., Robson, D., Li, J., Applebaum, S., Truong, T., Kesselman, C., Fraser, S. E., & Arnold, D. B. (2022a). *Mapping the Dynamic Synaptome*. https://synapse.isrd.isi.edu

Dempsey, W. P., Du, Z., Nadtochiy, A., Smith, C. D., Czajkowski, K., Andreev, A., Robson, D., Li, J., Applebaum, S., Truong, T., Kesselman, C., Fraser, S. E., & Arnold, D. B. (2022b). Regional synapse gain & loss accompany memory formation in larval zebrafish. *Proceedings of the National Academy of Sciences*.

Deng, Y., Kesselman, C., Sen, S., & Xu, J. (2019). Computational Operations Research Exchange (Core): A Cyber-Infrastructure for Analytics. 2019 Winter Simulation Conference (WSC),

Deutsch, P. (1996). Rfc1951: Deflate compressed data format specification version 1.3. In: RFC Editor.

Editorial, N. (2017). On data availability, reproducibility and reuse. *Nat Cell Biol*, *19*(4), 259. https://doi.org/10.1038/ncb3506

Errington, T. M., Alexandria Denis, Nicole Perfito, Iorns, E., & Nosek, B. A. (2021). Reproducibility in Cancer Biology: Challenges for assessing replicability in preclinical cancer biology. *eLife*. https://doi.org/DOI: 10.7554/eLife.67995

Feynman, R. P. (1974). Cargo cult science. *Engineering and Science*, *37*(7), 10-13.

Foster, E. D., & Deardorff, A. (2017). Open science framework (OSF). *Journal of the Medical Library Association: JMLA*, *105*(2), 203.

Gibbs, S., Moore, K., Steel, G., & McKinnon, A. (2017). The Dunning-Kruger Effect in a workplace computing setting. *Computers in Human Behavior*, *72*, 589-595. https://doi.org/https://doi.org/10.1016/j.chb.2016.12.084

Goldberg, I. G., Allan, C., Burel, J.-M., Creager, D., Falconi, A., Hochheiser, H., Johnston, J., Mellen, J., Sorger, P. K., & Swedlow, J. R. (2005). The Open Microscopy Environment (OME) Data Model and XML file: open tools for informatics and quantitative analysis in biological imaging. *Genome biology*, *6*(5), 1-13.

Gross, G. G., Junge, J. A., Mora, R. J., Kwon, H.-B., Olson, C. A., Takahashi, T. T., Liman, E. R., Ellis-Davies, G. C., McGee, A. W., & Sabatini, B. L. (2013). Recombinant probes for visualizing endogenous synaptic proteins in living neurons. *Neuron*, *78*(6), 971-985.

Halevy, A., Korn, F., Noy, N. F., Olston, C., Polyzotis, N., Roy, S., & Whang, S. E. (2016). Goods: Organizing google's datasets. Proceedings of the 2016 International Conference on Management of Data,

Harding, S. D., Armit, C., Armstrong, J., Brennan, J., Cheng, Y., Haggarty, B., Houghton, D., Lloyd-MacGilp, S., Pi, X., & Roochun, Y. (2011). The GUDMAP database−an online resource for genitourinary research. *Development*, *138*(13), 2845-2853.
Hauréau, B. (1852). *Histoire littéraire du Maine* (Vol. 1). Julien, Lanier et Ce.
Hopkin, M. (2006). Troublesome gene names get the boot. *Nature doi*, *10*.
Hsu, C.-N., Bandrowski, A. E., Gillespie, T. H., Udell, J., Lin, K.-W., Ozyurt, I. B., Grethe, J. S., & Martone, M. E. (2019). Comparing the use of research resource identifiers and natural language processing for citation of databases, software, and other digital artifacts. *Computing in Science & Engineering*, *22*(2), 22-32.
Humpel, C. (2018). Organotypic brain slice cultures. *Current Protocols in Immunology*, *123*(1), e59.
Jones, E. G. (1994). Santiago Ramón y cajal and the croonian lecture, march 1894. *Trends in neurosciences*, *17*(5), 190-192.
Kanare, H. M. (1985). *Writing the laboratory notebook*. ERIC.
Katz, D., Druskat, S., Cosden, I., Richmond, P., & Fouilloux, A. (2021, January 27, 2021). Introducing the International Council of RSE Associations. https://researchsoftware.org/2021/01/27/introducing-the-international-council-of-RSE-associations.htm
Kozlov, M. (2022). NIH issues a seismic mandate: share data publicly. *Nature*, *602*.
Lifanov, J., Linde-Domingo, J., & Wimber, M. (2021). Feature-specific reaction times reveal a semanticisation of memories over time and with repeated remembering. *Nature communications*, *12*(1), 1-10.
Madden, E., Scancella, J., & Adams, C. (2018). *The BagIt File Packaging Format (V1.0)* [RFC](RFC 8493). https://www.rfc-editor.org/info/rfc8493
Madduri, R., Chard, K., D'Arcy, M., Jung, S. C., Rodriguez, A., Sulakhe, D., Deutsch, E., Funk, C., Heavner, B., Richards, M., Shannon, P., Glusman, G., Price, N., Kesselman, C., & Foster, I. (2019). Reproducible big data science: A case study in continuous FAIRness. *PloS one*, *14*(4), e0213013. https://doi.org/10.1371/journal.pone.0213013
Martone, M., & Stall, S. *NIH Workshop on the Role of Generalist Repositories to Enhance Data Discoverability and Reuse: Workshop Summary*. https://datascience.nih.gov/data-ecosystem/NIH-data-repository-workshop-summary
Noy, N. F., & Klein, M. (2004). Ontology evolution: Not the same as schema evolution. *Knowledge and information systems*, *6*(4), 428-440.
Oldenburg, H. (1665). Introduction. *Philosophical Transactions*, *1*(1-2).
Paskin, N. (2010). Digital object identifier (DOI®) system. *Encyclopedia of library and information sciences*, *3*, 1586-1592.
*Python Package Index - PyPI*. Python Software Foundation. https://pypi.org
Roche, D. G., Kruuk, L. E., Lanfear, R., & Binning, S. A. (2015). Public data archiving in ecology and evolution: how well are we doing? *PLoS biology*, *13*(11), e1002295.
Samuels, B. D., Aho, R., Brinkley, J. F., Bugacov, A., Feingold, E., Fisher, S., Gonzalez-Reiche, A. S., Hacia, J. G., Hallgrimsson, B., & Hansen, K. (2020). FaceBase 3: analytical tools and FAIR resources for craniofacial and dental research. *Development*, *147*(18), dev191213.
Sandve, G. K., Nekrutenko, A., Taylor, J., & Hovig, E. (2013). Ten simple rules for reproducible computational research. *PLoS Comput Biol*, *9*(10), e1003285. https://doi.org/10.1371/journal.pcbi.1003285
Schneier, B., & Schneier, B. (2003). *Beyond fear: Thinking sensibly about security in an uncertain world* (Vol. 10). Springer.

Schuler, R., Czajkowski, K., D'Arcy, M., Tangmunarunkit, H., & Kesselman, C. (2020). Towards co-evolution of data-centric ecosystems. 32nd International Conference on Scientific and Statistical Database Management,

Schuler, R., & Kesselman, C. (2021). CHiSEL: a user-oriented framework for simplifing database evolution. *Distributed and Parallel Databases*, *39*(2), 483-543.

Schuler, R., Kesselman, C., & Czajkowski, K. (2016). Accelerating data-driven discovery with scientific asset management. 2016 IEEE 12th International Conference on e-Science (e-Science),

Schuler, R., Samuels, B. D., Bugacov, A., Williams, C., Ho, T.-V., Hacia, J. G., Pearlman, L., Iwata, J., Zhao, Z., Kesselman, C., & Chai, Y. (2022). FaceBase: A Community-Driven Hub for Data Intensive Research in Dental and Craniofacial Development and Dysmorphology. *Journal of Dental Research*, *Special Issue on Data-Driven Analytics for Dental, Oral, and Craniofacial Health Care*.

Singla, J., McClary, K. M., White, K. L., Alber, F., Sali, A., & Stevens, R. C. (2018). Opportunities and Challenges in Building a Spatiotemporal Multi-scale Model of the Human Pancreatic β Cell. *Cell*, *173*(1), 11-19. https://doi.org/10.1016/j.cell.2018.03.014

Smith, M., Barton, M., Bass, M., Branschofsky, M., McClellan, G., Stuve, D., Tansley, R., & Walker, J. H. (2003). DSpace: An open source dynamic digital repository.

Soiland-Reyes, S., Sefton, P., Crosas, M., Castro, L. J., Coppens, F., Fernández, J. M., Garijo, D., Grüning, B., La Rosa, M., Leo, S., Ó Carragáin, E., Portier, M., Trisovic, A., Community, R. O.-C., Groth, P., & Goble, C. (2022). Packaging research artefacts with RO-Crate. *Data Science*, *Preprint*, 1-42. https://doi.org/10.3233/DS-210053

Stonebraker, M., Deng, D., & Brodie, M. L. (2016). Database decay and how to avoid it. 2016 IEEE International Conference on Big Data (Big Data),

Tansley, R., Smith, M., & Walker, J. H. (2005). The DSpace open source digital asset management system: challenges and opportunities. International Conference on Theory and Practice of Digital Libraries,

Tedersoo, L., Kungas, R., Oras, E., Koster, K., Eenmaa, H., Leijen, A., Pedaste, M., Raju, M., Astapova, A., Lukner, H., Kogermann, K., & Sepp, T. (2021). Data sharing practices and data availability upon request differ across scientific disciplines. *Sci Data*, *8*(1), 192. https://doi.org/10.1038/s41597-021-00981-0

Vallat, B., Webb, B., Fayazi, M., Voinea, S., Tangmunarunkit, H., Ganesan, S. J., Lawson, C. L., Westbrook, J. D., Kesselman, C., & Sali, A. (2021). New system for archiving integrative structures. *Acta Crystallographica Section D: Structural Biology*, *77*(12).

Walsh, P., & Pollock, R. (2021, October 5, 2021). *Table Schema*. frictionless.io. https://specs.frictionlessdata.io/table-schema/

Wilkinson, M. D., Dumontier, M., Aalbersberg, I. J., Appleton, G., Axton, M., Baak, A., Blomberg, N., Boiten, J. W., da Silva Santos, L. B., Bourne, P. E., Bouwman, J., Brookes, A. J., Clark, T., Crosas, M., Dillo, I., Dumon, O., Edmunds, S., Evelo, C. T., Finkers, R., . . . Mons, B. (2016). The FAIR Guiding Principles for scientific data management and stewardship. *Sci Data*, *3*(1), 160018. https://doi.org/10.1038/sdata.2016.18

Wing, J. M. (2019). The Data Life Cycle. *Harvard Data Science Review*. https://doi.org/10.1162/99608f92.e26845b4

Ziemann, M., Eren, Y., & El-Osta, A. (2016). Gene name errors are widespread in the scientific literature. *Genome Biol*, *17*(1), 177. https://doi.org/10.1186/s13059-016-1044-7